\documentclass{article}
\usepackage{fullpage}
\usepackage{graphicx}
\usepackage{amsmath}
\usepackage{amssymb}
\title{The Cornell Potential from General Geometries in AdS/QCD}
\date{}
\renewcommand{\vec}[1]{\mbox{\boldmath$ #1 $}}

\begin{document}
\bibliographystyle{utphys}
\newcommand{\msbar}{\ensuremath{\overline{\text{MS}}}}
\newcommand{\DIS}{\ensuremath{\text{DIS}}}
\newcommand{\abar}{\ensuremath{\bar{\alpha}_S}}
\newcommand{\bb}{\ensuremath{\bar{\beta}_0}}
\newcommand{\rc}{\ensuremath{r_{\text{cut}}}}
\newcommand{\Nd}{\ensuremath{N_{\text{d.o.f.}}}}
\setlength{\parindent}{0pt}

\titlepage
\begin{flushright}
NIKHEF/2007-001 \\
\end{flushright}

\vspace*{0.5cm}

\begin{center}
{\Large \bf The Cornell Potential from General Geometries in AdS/QCD}

\vspace*{1cm}
\textsc{C.D. White$^{a,}$\footnote{cwhite@nikhef.nl} } \\

\vspace*{0.5cm} $^a$ NIKHEF, Kruislaan 409, 1098 SJ Amsterdam, The Netherlands\\

\end{center}

\vspace*{0.5cm}

\begin{abstract}
We consider the heavy quark-antiquark potential in the AdS / QCD correspondence, focusing in particular on a recently calculated AdS-like metric deformed by back-reaction effects. We find that tuning the long-distance behaviour of the potential leads to a discrepancy at small distances, and discuss how to better constrain AdS / QCD geometries. A systematic comparison of various geometries is presented, based on goodness of fit to lattice data in the quenched approximation. The back-reacted geometry is seen to be phenomenologically favoured over an alternative geometry with the same number of parameters, although it does not perform as well as some other geometries.
\end{abstract}

\vspace*{0.5cm}

\section{Introduction}
Regardless of whether string theory provides a fundamental description of nature, there is mounting evidence in favour of its being a useful tool for understanding strongly coupled gauge theories. The AdS / CFT correspondence of Maldacena \cite{Maldacena} conjectures a mathematical equivalence between the low energy limit of type IIB string theory on AdS$_5\otimes$S$^5$ and a conformal field theory, ${\cal N}=4$ $U(N)$ Super-Yang-Mills (SYM) for large $N$, on the boundary of this space. Since this initial conjecture, various generalisations of the duality principle have been considered involving other gauge theories and their proposed gravity duals. Crucially, a weakly coupled string theory corresponds to a strongly coupled gauge theory, and it is this apparent fact that generates much phenomenological interest in such correspondences. The hope is that a geometry might be found whose dual field theory mimics Quantum Chromodynamics (QCD), thus facilitating the use of string theory to ascertain the properties of strong interactions. There are two main approaches. Firstly, one can develop consistent string theories and try to ascertain the properties of their field theory duals, with the hope of edging closer to QCD. Such theories are ten dimensional and must then be compactified with consequent ambiguities. A second more phenomenological approach is to try to guess a (string-inspired) effective field theory in five-dimensional geometry whose dual (4-dimensional) theory then has QCD-like properties. This is known as the AdS / QCD approach by analogy with Maldacena's conjecture, although the term is slightly misleading due to the fact that deformed Anti-de-Sitter spaces are usually considered. There are then no ambiguities due to choice of compactification manifold, but the choice of 5-dimensional geometry one starts with is itself undetermined. Thus, it is important to classify the properties of various geometries and to examine the constraints which may be imposed to rule out various alternative geometries. \\

In order to determine whether or not a particular geometry (which may or may not have a specific theoretical motivation) approximates QCD well, it is useful to consider more than one geometry and compare results coming from the AdS / QCD correspondence in a systematic manner. The aim of this paper is to undertake such an investigation, using the static interquark potential as an observable to constrain the theory. The usefulness of this quantity stems from the fact that one can easily separate the regime where the semi-classical string calculation is expected to hold (large distances) from the regime where the AdS / QCD approach is expected to break down (small distances). \\

A five-dimensional holographic hadron model was introduced in \cite{Erlich}. A class of models exhibiting linear confinement was examined in \cite{Karch,Andreev1}. In \cite{Andreev2} the heavy quark-antiquark potential was considered in a particular geometry belonging to this class, and found to be consistent with the {\it Cornell potential} \cite{Eichten}:
\begin{equation}
V(r)=V_0+\kappa r-\frac{e}{r}+\frac{f}{r^2},
\label{Cornell}
\end{equation}
where the coefficients can be found by fitting e.g. to Charmonium spectra. The original Cornell potential did not include the term in $r^{-2}$, which strictly speaking is not physical. However, such a term is often introduced when parameterising lattice determinations of the potential, in order to enhance the fit to data by modelling running coupling effects etc. For a review of lattice measurements of the potential, see \cite{Bali}. Absence of stringy corrections in the AdS / QCD calculation corresponds to a lack of quark dynamics. Thus in comparing the calculated potential with lattice results, one should use quenched data in which the quarks are not dynamical.\\

It has been shown previously that Cornell-like behaviour arises from the AdS / CFT correspondence subject to quite general conditions on the geometry \cite{Kinar}. In this paper we consider various geometries in the five-dimensional framework, including the recently calculated geometry of \cite{Shock}, in which the back-reaction of the 5-dimensional quark condensate fields are used to deform an initial AdS space. It is hoped that such a geometry better models the physics on the field theory side of the correspondence, which motivates a comparison with other possible geometries. We use a $\chi^2$ goodness of fit criterion to compare the potentials with lattice data. It is found that no geometry is able to describe the whole range of lattice data for the potential (which is to be expected), but that certain geometries are better able to describe the moderate and large distance data than the back-reacted geometry of \cite{Shock}. However, this latter geometry seems to encompass more physics than a similar geometry with the same number of variable parameters. A geometry based on a simple quadratic warp factor \cite{Andreev2} appears by inspection to agree most closely with the data, but this conclusion is argued to be somewhat misleading. \\

The paper is laid out as follows. In section \ref{Potential} we recall how to calculate the heavy quark-antiquark potential from a string world-sheet in 5-dimensions spanning a Wilson line on the 4-dimensional boundary \cite{Maldacena2,Andreev2}. In section 3 we present explicit results for the potential using the back-reacted geometry of \cite{Shock}, using the parameters fixed as in that paper and the procedure of \cite{Andreev2}. A discrepancy is found between the potentials at small $r$ values, and in section 4 we consider a more general and systematic procedure for constraining AdS / QCD geometries. Section 5 contains a discussion of the results, and conclusions.
\section{The Heavy Quark-Antiquark Potential}
\label{Potential}
The short-distance $r^{-1}$ behaviour of the interquark potential follows from the asymptotic AdS behaviour of the bulk space \cite{Maldacena}, and linear confinement at large distances was demonstrated on quite general grounds in \cite{Kinar}. For completeness and to set up our notation, we recall here how to calculate the interquark potential. \\

In field theory, the potential between two sources of the gauge field is calculated from the expectation value of a Wilson loop $W({\cal C})$ connecting them in space-time. In fact, one has:
\begin{equation}
<W({\cal C})>\propto e^{-T V(r)}
\label{wilson}
\end{equation}
where the limit $T\rightarrow\infty$ is understood and $V(r)$ is the separation-dependent potential. According to the AdS / CFT correspondence, the expectation value of a Wilson loop in the boundary conformal field theory is related to the extremal area of a string world sheet in the higher-dimensional theory, which spans the curve ${\cal C}$ \cite{Maldacena2}. One has:
\begin{equation}
<W({\cal C})>\propto e^{-A},
\label{wilson2}
\end{equation}
where $A$ is the area of the string worldsheet. Combining this with equation (\ref{wilson}), the potential (up to a constant) is given by:
\begin{equation}
V(r)=\frac{1}{2\pi\alpha' T}\int d^2\xi\sqrt{\text{det}g_{nm}\partial_\alpha X^n\partial_\beta X^m},
\label{pot}
\end{equation}
where the Nambu-Goto action for the string worldsheet has been used, and it is understood that this worldsheet has extremal area. Here $\alpha'$ is the string tension, $g_{nm}$ the space-time metric, $X^n(\vec{\xi})$ are the space-time coordinates and $\vec{\xi}=(\xi_0,\xi_1)$ are the worldsheet coordinates. One can now find the form of the potential by assuming a given background metric $g_{nm}(X)$, and it is therefore at this point that the model-dependence enters. This method was applied in the framework of the AdS / QCD approach in \cite{Andreev2,Braga}. Here we assume a general form of the metric:
\begin{equation}
ds^2=f(z)\,dx^\mu dx_\mu+\frac{dz^2}{z^2},
\label{metric}
\end{equation}
where $z$ is the fifth dimension and $x^\mu$ the 4-dimensional coordinates. If $f(z)=z^{-2}$, this gives a Euclidean AdS space. The possible form of the deformation function $f(z)$ will be considered later. Substituting equation (\ref{metric}) into equation (\ref{pot}) and choosing $\xi_0=t$ and $\xi_1=x$ (possible because of reparameterisation invariance of the worldsheet), one obtains following the method of \cite{Andreev2}:
\begin{equation}
r(z_0)=\int_{-\frac{r}{2}}^{\frac{r}{2}}dx=2\int_0^{z_0}dz\frac{1}{\sqrt{\tilde{f}}}\left[\frac{z_0^4\tilde{f}^2}{z^4\tilde{f}_0^2}-1\right]^{-\frac{1}{2}}
\label{r}
\end{equation}
and:
\begin{equation}
V_R(z_0)=\frac{g}{\pi}\int_0^{z_0}\frac{dz}{z^2}\left[\sqrt{\tilde{f}}\left(1-\frac{z^4\tilde{f}_0^2}{z_0^4\tilde{f}^2}\right)^{-\frac{1}{2}}\right],
\label{V0}
\end{equation}
where $z_0$ is the value of $z$ at $x=0$, $r(z_0)$ the interquark separation and $g=1/\alpha'$. We have also introduced $\tilde{f}(z)=f(z)/z^2$, which aids numerical convergence of the integrals. The potential given by equation (\ref{V0}) diverges and must be regularised \cite{Maldacena2}, as seen by the fact that $\tilde{f}(z)\rightarrow 1$ as $z\rightarrow 0$ from the asymptotic AdS behaviour of the metric. Replacing the lower limit of the integral in equation \ref{V0} by $z=\delta$, one finds a regularised energy: 
\begin{equation}
V_R(z_0)\equiv\lim_{\delta\rightarrow 0}\left[V(z_0,\delta)-\frac{g}{\pi}\frac{1}{\delta}\right]=\frac{g}{\pi}\left\{-\frac{1}{z_0}+\int_0^{z_0}\frac{dz}{z^2}\left[\sqrt{\tilde{f}}\left(1-\frac{z^4\tilde{f}_0^2}{z_0^4\tilde{f}^2}\right)^{-\frac{1}{2}}-1\right]\right\}.
\label{VR}
\end{equation}
Under general conditions for $f(z)$ the potential will be Cornell-like \cite{Maldacena2,Kinar}, with asymptotic limits:
\begin{align}
V_R(r)&\rightarrow -\frac{g}{2\pi\rho^2 r},\quad r\rightarrow0;\label{ER}\\
&\rightarrow \frac{g\tilde{f}_0^*}{2\pi (z_0^*)^2}r,\quad r\rightarrow\infty,\label{VRasympt}
\end{align}
where $\rho=\Gamma^2(1/4)/(2\pi)^{3/2}$ and $\{z_0^*,\,\tilde{f}_0^*\}$ are the values of $z_0$ and $f(z_0)$ at which the interquark separation diverges to infinity. This will occur according to the criterion:
\begin{equation}
\frac{d}{dz}\left[z_0^4\tilde{f}^2\right]_{z_0^*}=\frac{d}{dz}\left[z^4\tilde{f}_0^2\right]_{z_0^*},
\label{crit}
\end{equation}
which gives the equation for $z_0^*$:
\begin{equation}
z_0^*\left.\frac{d\tilde{f}}{dz}\right|_{z_0^*}=2\tilde{f}_0^*,
\label{z0*}
\end{equation}
In general one is only able to solve this equation numerically for $z_0^*$, although we will see in the next section that an analytical solution is possible for the back-reacted geometry of \cite{Shock}. It is furthermore useful to introduce the following ansatz for the deformation function $f(z)$:
\begin{equation}
f(z)=e^{2A(z)}.
\label{ansatz}
\end{equation}
\section{Results in a Back Reacted Geometry}
\label{back}
The holographic hadron model of \cite{Erlich} features a five dimensional field dual to the 4-dimensional bilinear quark operator $q\bar{q}$ permeating a fixed 5-dimensional (AdS) geometry. This is an approximation which ignores the fact that the action for the 5-dimensional field theory in a curved space-time couples the quark field to the metric. Thus, the presence of a bulk field will itself deform the AdS geometry, leading to the deformation factor $f(z)$ discussed above. This effect was taken into account in \cite{Shock}, where a differential equation is given for the ``warp factor'' $A(z)$ of equation (\ref{ansatz}) in terms of the bulk scalar field \footnote{Note that the dilaton $\phi$ is assumed constant in this analysis, consistent with the semi-classical approximation.}. One solves this equation subject to a suitable ansatz for this field, and more than one example is given in \cite{Shock}. We use as our example the following back-reacted warp-factor (denoted ``Model 1'' in \cite{Shock}):
\begin{equation}
A(z)=-\log{z}+\frac{m_q^2}{48}z^2-\frac{m_q\sigma}{32}z^4+\frac{\sigma^2}{48}z^6,
\label{A(z)}
\end{equation}
coming from the following ansatz \cite{Erlich} for the bulk field $X$ dual to the bilinear quark operator\footnote{This is not to be confused with the string space-time coordinates introduced earlier.}:
\begin{equation}
X=\frac{m_q}{2}z+\frac{\sigma}{2}z^3.
\label{phi}
\end{equation}
Care is needed to obtain equation (\ref{A(z)}) from the result given in \cite{Shock} due to the use of a Euclidean signature here. This technicality is explained in appendix \ref{append}. Using equations (\ref{z0*}, \ref{ansatz}), we can find the value $z_0=z_0^*$ at which the interquark separation diverges. It is given by the equation:
\begin{equation}
\frac{\sigma^2}{8}(z_0^*)^6-\frac{m_q\sigma}{8}(z_0^*)^4+\frac{m_q^2}{24}(z_0^*)^2-1=0,
\label{cubic}
\end{equation}
which is a cubic equation for $(z_0^*)^2$, whose real solution for general $m_q$ and $\sigma$ is:
\begin{equation}
z_0^*(m_q,\sigma)=\left[\frac{m_q\sigma+\sqrt[3]{216\sigma^4-m_q^3\sigma^3}}{3\sigma^2}\right]^{\frac{1}{2}}.
\label{z0gen}
\end{equation}
In \cite{Shock} the values of $m_q$ and $\sigma$ were constrained by considering the number of quark colours, meson masses and the pion decay constant, thus leading to the values $m_q=2.4\text{MeV},\,\sigma=(326\text{MeV})^3$.
Then one has:
\begin{equation}
z_0^*=4.341\,\text{GeV}^{-1}, \quad \tilde{f}_0^*=1.395
\label{z0val}
\end{equation}
for the extremal values of the parameter $z_0$ and the deformation function. From equation (\ref{VRasympt}), one has for these values of $m_q$ and $\sigma$:
\begin{equation}
V_R\rightarrow (0.01178g)r,\quad r\rightarrow\infty
\label{VRval}
\end{equation}
where $g$ has yet to be fixed. In \cite{Andreev2}, this parameter is fixed from the Cornell potential. The lattice parameterisation of \cite{data1} gives $\kappa=0.183$ in equation (\ref{Cornell}). Adopting this procedure here, one finds $g=15.55$. This is much larger than the value $g=0.94$ given in \cite{Andreev2} (albeit for a slightly different geometry), this difference mostly arising from the parameters of the warp factor in the geometry. If one fixes the values of $m_q$ and $\sigma$ from \cite{Shock}, one has no choice but to fix $g$ in this way in order to obtain the correct long-distance behaviour. The potential can be directly compared with lattice data in the quenched approximation, where dynamics of the quarks is suppressed. We use the data\footnote{We are grateful to Gunnar Bali for providing files of the data from \cite{data1,data2}. Further precise lattice data exists at low $r$ values \cite{Weisz,Koma}, but this will not affect the results presented in this paper.} of \cite{data1,data2,data3}, shown in figure 3 as a function of $r/r_0$, where $r_0=0.5\text{fm}\equiv 2.5\text{GeV}$, and with a constant added such that $V(r_0)=0$.
\begin{figure}
\begin{center}
\scalebox{0.7}{\includegraphics{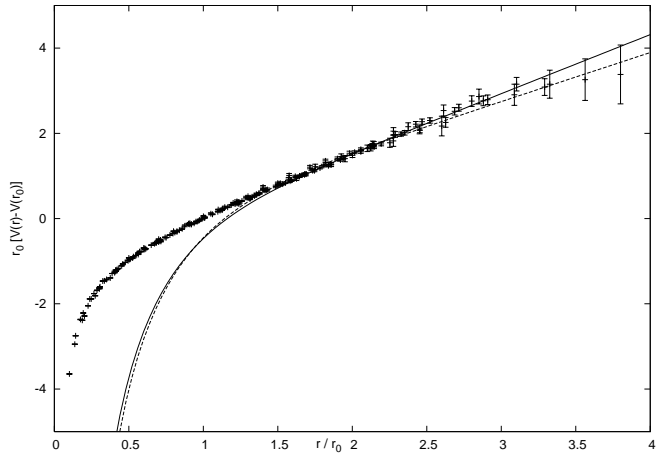}}
\caption{The potential $V_R(r)$ arising from the geometry (\ref{A(z)}) with parameters as discussed in the text (solid), alongside the quenched lattice data of \cite{data1,data2,data3}. The constant term has been adjusted. Also shown is the Cornell potential obtained from fitting the parameters (dashed), as described in section \ref{constrain}.}
\label{energies}
\end{center}
\end{figure}
The constant term of the AdS potential is also adjusted so that equality is reached with the lattice potential at asymptotically large distances, in order to more easily facilitate a comparison. One sees a marked deviation between the two potentials away from the asymptotically long-distance regime. Indeed the value of $g$ obtained gives, via equation (\ref{ER}), the short-distance behaviour:
\begin{equation}
V_R(r)\rightarrow-\frac{3.55}{r},\quad r\rightarrow 0
\label{VR0val}
\end{equation}
in marked contrast to the lattice potential used to parameterise the data in \cite{data1}, which has $e=0.321$ in equation (\ref{Cornell}). This difference has arisen from fixing $m_q$ and $\sigma$ according to \cite{Shock}, then using the long-distance behaviour of the Cornell potential to fix the only remaining free parameter $g$. There are then no free parameters left which influence the short-distance behaviour - $g$ is the only parameter governing this regime. Although one does not expect the semi-classical theory used here to give a correct description at short distances, the possibility exists of using the parameters in the deformation function $f(z)$ to tune the long-distance behaviour of the potential. Then $g$ can be used so as to extend agreement between the AdS / QCD and Cornell potentials into the moderate $r$ region, and the parameters $m_q$ and $\sigma$ varied to tune the long-distance behaviour. This exercise can be repeated for different candidate geometries, and the results used to constrain geometrical parameters, and also compare the phenomenological validity of the various models. This is the subject of the following section.
\section{Constraining AdS / QCD Models}
\label{constrain}
Figure \ref{energies} compares the potential obtained by fixing the parameters according to \cite{Shock} with lattice data. This gives no quantitative indication of how good the agreement is. To measure the goodness of fit of the theoretical potential to the data, one may define the standard $\chi^2$ variable:
\begin{equation}
\chi^2=\sum_{i=1}^{N_{\text{data}}}\frac{(V_i^{\text{AdS/QCD}}-V_i^{\text{latt.}})^2}{\sigma_i^2},
\label{chi2}
\end{equation}
where the sum is over the $N_{\text{data}}$ data points, and $\sigma_i$ is the absolute error associated with the $i^{\text{th}}$ point. We also introduce the number of degrees of freedom:
\begin{equation}
N_{\text{d.o.f}}=N_{\text{data}}-N_{\text{param}},
\label{ndof}
\end{equation}
where $N_{\text{param}}$ is the number of free parameters associated with the geometry. When minimised, a $\chi^2/\Nd$ of approximately 1 denotes a good fit, and here a problem occurs in that it is not possible to achieve a good fit to all data points using the potential of equations (\ref{r}, \ref{VR}). This is to be expected given the discrepancy at low $r$ values noted in the previous section after fixing parameters at high $r$. One of course expects the AdS / QCD calculation to break down away from asymptotically large $r$ values, due to a multitude of reasons. Firstly, there are stringy corrections to the semi-classical approximation employed here. Secondly, the type of five-dimensional description used in this paper must in principle arise from compactification of a 10-dimensional string theory, none of which physics is well-known. The very small $r$ behaviour is accessible in the field theory description, and one expects modifications of the na\"{i}ve $r^{-1}$ behaviour due to the running of the coupling. \\

\begin{figure}
\begin{center}
\scalebox{0.7}{\includegraphics{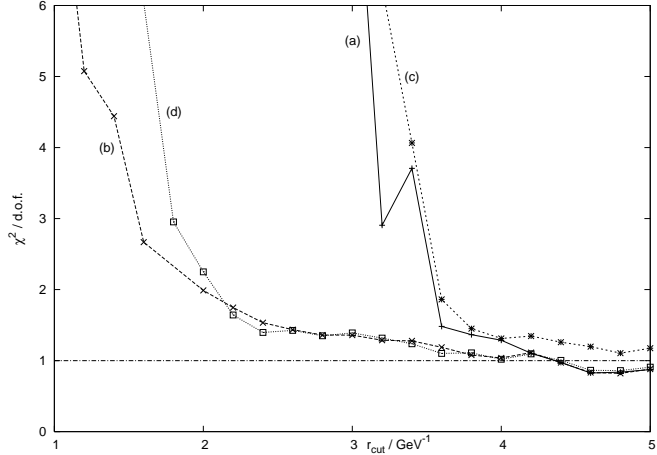}}
\caption{Variation of the minimum $\chi^2/\Nd$ with the data cut $\rc$ for (a) the back-reacted geometry of \cite{Shock}; (b) the quadratic warp factor geometry of \cite{Andreev2}; (c) the geometry denoted III in table \ref{geometries} (d) the geometry denoted IV in table \ref{geometries}.}
\label{chiplot}
\end{center}
\end{figure}

With this level of ignorance, the best one can do is to impose a cut $\rc$ on the lattice data, excluding all values satisfying $r<\rc$. Physically, this corresponds to tuning the AdS / QCD geometry parameters using data only in the regime where the calculation is expected to hold. For each value of $r_{cut}$, there is a corresponding minimum value of $\chi^2$ that can be achieved by varying the geometry. One then gains some idea of where the string calculation breaks down by looking at how the minimum goodness of fit per degree of freedom, $\chi^2(r_{cut})/\Nd$, varies with $\rc$. This is shown in figure \ref{chiplot} for the back-reacted geometry considered in the previous section, marked as (a) in the figure.
One sees that one can fit the theoretical potential to the lattice data well for $r$ values down to approximately 4.4 GeV$^{-1}$, before the goodness of fit per degree of freedom rises sharply as more data is added. As well as the parameters of the warp factor, the coupling $g$ is also allowed to vary, together with a constant term added to the potential as this has no effect on dynamics.\\

The potential obtained by fitting for a value of $r_{cut}$ such that the minimum $\chi^2/\Nd\simeq 1$ is shown in figure \ref{energies}. One sees that the fit has not improved the description at small distances - indeed, the fitted potential actually lies slightly further from the lattice data at low $r$. Instead the slope at large $r$ has varied in order to increase the goodness of fit. This is a clear indication that the string calculation breaks down at moderate $r$ values. Of course, there is no guarantee that this geometry corresponds to QCD. In the above analysis one is, after all, fitting a function known to be asymptotically linear with some data which also displays such behaviour. It is then only natural that at some point the theory and data must diverge, and it is not clear that this is really due to the breakdown of the AdS / QCD approximation. However, further insight is gained by repeating the above analysis with different geometries. \\

We use for comparison the metric of \cite{Andreev2} which has a warp factor containing a quadratic term in the exponent; also a geometry with a two-parameter exponent involving terms in $z^2$ and $z^4$; finally a three parameter exponent involving terms up to and including $z^6$. The different metrics are summarised in table 1, with the latter two denoted by III and IV.
\begin{table}[h]
\begin{center}
\begin{tabular}{c|c|c}
Geometry & $\bar{A}(z)$ &Parameters\\
\hline
Back Reacted \cite{Shock}& $\frac{m_q^2}{48}z^2-\frac{m_q\sigma}{32}z^4+\frac{\sigma^2}{48}z^6$& $m_q$, $\sigma$, $g$, const.\\
Andreev et. al. \cite{Andreev2} & $\frac{c}{4}z^2$&$c$, $g$, const.\\
III&$k_1z^2+k_2z^4$&$k_1$, $k_2$, $g$, const.\\
IV&$k_1z^2+k_2z^4+k_3z^6$&$k_1$, $k_2$, $k_3$, $g$, const.\\
\end{tabular}
\label{geometries}
\caption{The geometries used to compare AdS / QCD predictions with the lattice data, where $\bar{A}(z)=A(z)+\log(z)$ denotes the polynomial terms in the exponent of the warp factor. The coupling $g$ is also allowed to vary, as is an arbitrary constant term added to the potential.} 
\end{center}
\end{table}
The first thing to note from figure \ref{chiplot} is that if one takes the criterion $\chi^2/\Nd<1$ very seriously, it is not possible to distinguish which geometry if any best describes the lattice data. Each satisfies this goodness of fit only for $\rc\gtrsim 4.4\text{GeV}^{-1}$. However it is clear that whilst the $\chi^2$ increases sharply for the back-reacted and III geometries as $\rc$ decreases below this value, it is still possible to fit the data well using the quadratic warp factor geometry of \cite{Andreev2}. This is despite the fact that the back-reacted geometry has an additional parameter. Nevertheless, the back-reacted geometry does outperform geometry III, which has the same number of parameters but not the same dependence on $z$ in the exponent of the warp factor. Thus, it seems that the back-reacted geometry is indeed to some extent modelling the physics more correctly than a na\"{i}ve expansion in $z$ containing the same number of parameters.\\

An improved fit is obtained by adding an additional parameter, as in geometry IV. Here the warp factor has the same polynomial as the back-reacted geometry, but the coefficients are allowed to vary independently. From figure \ref{chiplot}, one sees that this improves the range over which the lattice data can be fit and that this range roughly agrees with the result of the quadratic warp factor. \\

For each geometry, one obtains from figure \ref{chiplot} an approximate  value $\rc=\rc^*$ at which the $\chi^2$ per degree of freedom is approximately unity. The `best fit' parameter values are then the values arising from fits with these cuts. These are collected in table \ref{bestfit}. 
\begin{table}
\begin{center}
\begin{tabular}{c|c|c|c}
Geometry&$\rc^*$&Parameters&Best Fit\\
\hline
Back Reacted&4.35&$m_q$&0.0484\\
&&$\sigma$&0.0550\\
&&const.&0.244\\
&&$g$&14.2\\
\hline
Andreev et. al.&4.0&$c$&0.888\\
&&const.&-0.968\\
&&g&1.23\\
\hline
III&6.1&$k_1$&0.000598\\
&&$k_2$&0.00500\\
&&const.&0.240\\
&&$g$&15.7\\
\hline
IV&4.4&$k_1$&0.00185\\
&&$k_2$&0.000500\\
&&$k_3$&0.00100\\
&&const.&0.252\\
&&$g$&5.46\\
\end{tabular}
\caption{`Best fit' values for the parameters of table \ref{geometries} obtained by cutting the lattice data in each case such that the minimum $\chi^2/\Nd\simeq 1$, and then fitting. All parameter units are in terms of GeV.}
\end{center}
\label{bestfit}
\end{table}
One sees that when the coefficient of $z^4$ is allowed to vary independently from the $z^6$ coefficient, there is no preference for negative values. This may explain some of the improvement in the fit with respect to the back-reacted geometry, in which negativity of the $z^4$ coefficient is fixed as a consequence of the Wick rotation from the Euclidean space of \cite{Shock} together with the interpretation of $m_q$ and $\sigma$ in terms of quark mass and vacuum condensate terms respectively. \\

\begin{figure}
\begin{center}
\scalebox{0.7}{\includegraphics{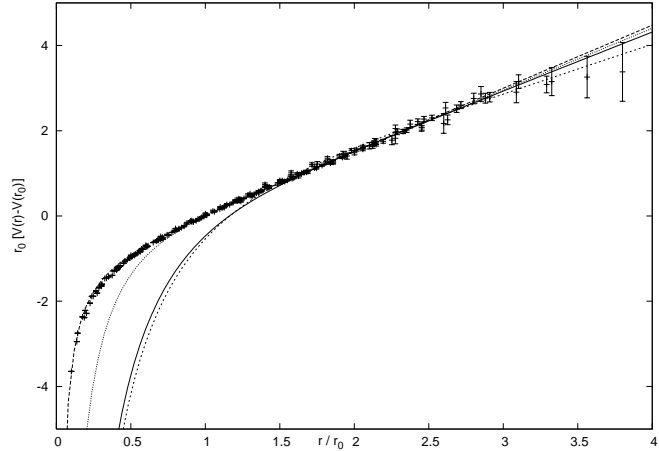}}
\caption{The static interquark potential arising from the `best fit' examples of each geometry: back-reacted geometry (solid); quadratic warp factor (long dashes); III (short dashes) and IV (dot dashed). The lattice data is shown with error bars.}
\label{potplot}
\end{center}
\end{figure}

The potentials arising from each of these `best fits' are shown in figure \ref{potplot}, alongside the lattice data.
All of the potentials produce the correct large $r$ behaviour as expected from the fits and the results of figure \ref{chiplot}. However, a striking feature of figure \ref{potplot} is the variation in the behaviour at small $r$. Geometry III is furthest from the lattice data at moderate and small $r$, which is consistent with the results of figure \ref{chiplot}. However, the potential from the back-reacted geometry does not fare much better. The geometry that appears to lie closest to all the data is the quadratic warp factor geometry, appearing to be even better than geometry IV which has two more parameters. However, this latter comparison is not meaningful as it is not valid to compare the potentials at values of $r$ at which the $\chi^2$ per degree of freedom is much greater than unity. Using $\rc=4\text{GeV}$ from the quadratic geometry (table \ref{bestfit}), this means one is not able to meaningfully compare the results for $r/r_0\lesssim1.6$ in figure \ref{potplot}. Thus even though by eye the quadratic geometry seems to fit the data over the whole range of $r$ very closely in figure \ref{potplot}, appearances are in this case deceptive. The fit is not good over this range when quantitatively measured using the $\chi^2$, and given that the quadratic geometry and geometry IV share a similar range over which they agree with the data at high $r$ one should regard the separation of the corresponding potential curves in figure \ref{potplot} as being some measure of the theoretical uncertainty in the calculation.
\section{Discussion}
In this paper we have considered the Cornell-like potential arising from the `Model I' geometry obtained by \cite{Shock} in their analysis of back-reaction effects. Cornell behaviour naturally arises consistent with previously found conditions \cite{Kinar}, and one can solve exactly for the asymptotic limits of the potential in terms of the parameters of this geometry. We examine the potential for the choice of parameters given in that paper. Once these are fixed, one can fix the only remaining parameter $g$ by matching to the long-distance behaviour of the Cornell potential. This gives a marked deviation between the AdS and Cornell potentials away from this regime, with a coefficient of the $r^{-1}$ behaviour at short distances that shows little agreement with the Cornell result. This leads to the more general question of how one should constrain AdS / QCD geometries, and we then give an example of a systematic procedure to test the validity of proposed models. \\

Two types of question arise. Firstly: of a given set of candidate geometries, which best approximates QCD? Secondly: if one knows that a particular type of geometry does indeed represent QCD, where does the AdS / QCD calculation break down? We have investigated both these questions by allowing the parameters of a number of geometries to vary simultaneously, and then quantifying the goodness of fit to lattice data using a standard $\chi^2$ minimisation. \\

The issue of where the string calculation is expected to break down can be investigated by varying the cut on the lattice data that is applied to filter out the region contaminated by unknown effects. A given geometry should not be trusted in the region where the lattice data increases the $\chi^2$ per data point to greater than approximately unity. For each of the potentials considered, this amounts to $r$ values less than a few GeV$^{-1}$. \\

Comparing the variation of $\chi^2/\Nd$ with $\rc$ for different geometries, as in figure \ref{chiplot}, gives some information on which geometries are closer to QCD. For example, the fact that the back-reacted geometry is better able to describe the lattice data (in the regime where we trust the calculation) than a geometry with the same number of parameters, is a clear indication that the physics is better described. This is encouraging, given that the back-reacted geometry has a better theoretical motivation. However, the simple quadratic geometry of \cite{Andreev2} gives better results. The fact that the resulting potential is closer to the lattice results at small $r$ (see figure \ref{potplot}) may nevertheless be misleading given that the $\chi^2$ per degree of freedom is no good in this regime, and there is a large theoretical uncertainty associated with the prediction. One can get closer to the data from the potential predicted by the back-reacted geometry by allowing more free parameters, with the same caveats at low $r$. Indeed, the fact that adding higher order terms in the warp factor brings the potential closer to the lattice data suggests that the good agreement from the quadratic warp factor may be a coincidence.\\

Another potential problem with the results presented here arises from the fact that we have only included the interquark potential in the fits, and it is not strictly correct to vary all the parameters of the geometry without also checking consistency with other observables that can be calculated from the theory. In principle, one should extend the $\chi^2$ definition of equation (\ref{chi2}) to include meson masses, decay constants etc. that also depend on the free parameters. However, the fact that the best fit parameters from the back-reacted geometry do not stray too far from the values originally fixed in \cite{Shock} means that one can be reasonably confident the additional constraints can be satisfied. This is not necessarily true for geometry IV, which has only been fit here to the lattice potential. Other observables are not as clean as the lattice potential in that it can be difficult to separate the regime where the AdS / QCD calculation is expected to hold from that where it breaks down.\\

\begin{figure}
\begin{center}
\scalebox{0.7}{\includegraphics{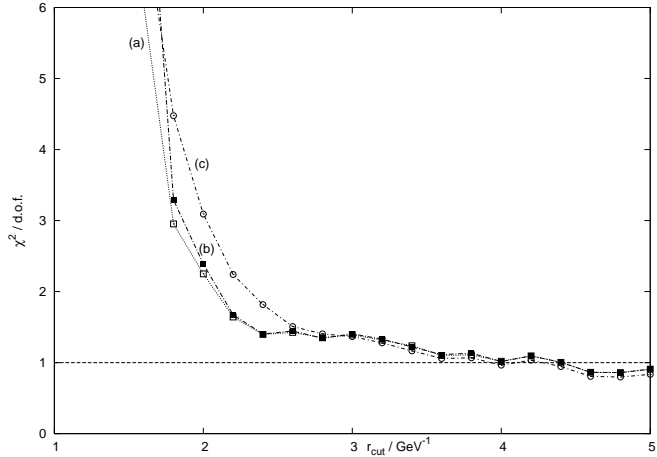}}
\caption{Variation of $\chi^2/\Nd$ with the data cut $\rc$ for (a) geometry IV with unconstrained parameters; (b) geometry IV with parameter $k_2$ constrained to be negative; (c) linear fit to the potential.}
\label{chiplot2}
\end{center}
\end{figure}

It may also be the case that varying the parameters over all possible values is unphysical. For example, in the back reacted geometry positivity of $m_q$ and $\sigma$ follows from their interpretation in terms of a heavy quark mass and vacuum condensate of the bilinear quark operator. This then leads to a negative $z^4$ term in the warp factor of the back-reacted geometry. No such constraint is present for the geometry IV, and it is therefore interesting to repeat the minimisation of section \ref{constrain} whilst ensuring that the coefficient $k_2$ of the $z^4$ term in the warp factor remains negative. The resulting variation of $\chi^2/\Nd$ with $\rc$ is shown in figure \ref{chiplot2}.
One sees that when $k_2$ is required to be negative, the resulting $\chi^2/\Nd$ does diverge for a slightly larger $\rc$ value but this is not a significant effect. \\

Also shown in figure \ref{chiplot2} is the goodness of fit of a straight line to the lattice data, which then gives some idea of where the linear regime breaks down. Comparing figures \ref{chiplot} and \ref{chiplot2} suggests that the back-reacted geometry begins to fail describing the lattice data even for $r$ values still in the linear regime. This is not the case for geometry IV and suggests that, whilst the back-reacted geometry seems to include more physics than the case of geometry III, the description is still incomplete. \\

Of course, the back-reacted geometry adopted here is not unique but relies upon a particular ansatz for the quark operator. A higher order expansion for this operator would result in higher order terms in the warp factor. It would be very interesting to see if these higher order terms improved the results shown here\footnote{Recently, the back-reacted geometry of \cite{Shock} has been extended to include the effect of the strange quark \cite{Wu}. This does not affect the results of this paper, as the nature of the polynomial in the warp factor is unchanged.}. One could also add extra terms to the warp factor which are unconstrained by back-reaction effects and compare the resulting fit with the back-reacted case. The resulting phenomenologically determined potential would then act as a constraint on various proposed theoretical models.\\

To conclude, we have used a $\chi^2$ goodness of fit measure to compare predictions for the static interquark potential from various geometries in the AdS / QCD correspondence with lattice data. It is possible to tell the difference between the predictions, with some clearly more favoured than others. A quadratic warp factor is favoured by the data, as is a potential having a higher order polynomial warp factor with parameters unconstrained by back-reaction. However, it still may be the case that higher order back-reacted geometries prove more successful in approximating QCD.  
\section{Acknowledgments}
CDW is supported by the Dutch Foundation for Fundamental Research of Matter (FOM). He would like to thank Heng-Yu Chen and Koenraad Schalm for conversations, Florian Gmeiner and Eric Laenen for numerous discussions and comments on the manuscript, and Oleg Andreev for correspondence regarding the use of a Euclidean metric.  
\appendix
\section{The Back-Reacted Geometry in Euclidean Space}
\label{append}
In \cite{Shock} back reaction effects are considered starting from the geometry:
\begin{equation}
ds^2=e^{-2\tilde{A}(y)}dx_\mu dx^\mu-dy^2,
\label{Shockmet}
\end{equation}
where we denote the warp factor with a tilde to avoid confusion with $A(z)$ as defined in this paper. Introducing $z=e^y$, one then has:
\begin{equation}
ds^2=e^{-2\tilde{A}(z)}dx_\mu dx^\mu-\frac{dz^2}{z^2}.
\label{Shockmet2}
\end{equation}
Note that instead of using a different symbol for the warp factor in terms of $z$, we clarify any ambiguity by denoting the arguments of this function explicitly. To obtain a Euclidean signature one can Wick rotate:
\begin{equation}
dx^\mu\rightarrow \imath dx^\mu,\quad z\rightarrow\imath z
\label{Wick}
\end{equation}
to obtain:
\begin{equation}
ds^2=\left[e^{-2\tilde{A}(\imath z)}\left(dt^2+d\vec{x}^2\right)-\frac{dz^2}{z^2}\right].
\label{Wickmet}
\end{equation}
The ``Model I'' warp factor of \cite{Shock} is given in Minkowski space by:
\begin{equation}
\tilde{A}(z)=\log(z)+\frac{m_q^2}{48}z^2+\frac{m_q\sigma}{32}z^4+\frac{\sigma^2}{48}z^6.
\label{AShock}
\end{equation}
Substituting this in equation (\ref{Wickmet}) gives:
\begin{equation}
|ds^2|=\left[\frac{1}{z^2}\exp\left(\frac{m_q^2}{24}z^2-\frac{m_q\sigma}{16}z^4+\frac{\sigma^2}{24}z^6\right)(dt^2+d\vec{x}^2)+\frac{dz^2}{z^2}\right].
\label{Wickmet2}
\end{equation}
We note that the requirement of a positive sign in a warp factor consisting only of $A(z)\sim z^2$ in Euclidean space was noted already in \cite{Andreev1} based on the need for a discrete spectrum of meson masses.  

\bibliography{refs.bib}
\end{document}